\documentstyle[multicol,prb,aps,epsfig]{revtex}
\begin{document}
\draft
\title{Resistance Noise Near to Electrical Breakdown:
\\Steady State of Random Networks as a Function of the Bias}

\author{C. Pennetta}

\address{INFM - National Nanotechnology Laboratory, 
and Dipartimento di Ingegneria dell'Innovazione, 
\\Universit\`a di Lecce, Italy, Via Arnesano, I-73100, Lecce, Italy,
\thanks{Corresponding authors e-mail: cecilia.pennetta@unile.it }}
%
%
\maketitle
\begin{abstract}
A short review is presented of a recently developed computational approach 
which allows the study of the resistance noise over the full range of bias 
values, from the linear regime up to electrical breakdown. Resistance noise 
is described in terms of two competing processes in a random resistor 
network. The two processes are thermally activated and driven by an 
electrical bias. In the linear regime, a scaling relation has been found 
between the relative variance of resistance fluctuations and the average 
resistance. The value of the critical exponent is significantly higher than 
that associated with 1/f noise. In the nonlinear regime, occurring when the 
bias overcomes the threshold value, the relative variance of resistance 
fluctuations scales with the bias. Two regions can be identified in this 
regime: a moderate bias region and a pre-breakdown one. In the first region, 
the scaling exponent has been found independent of the values of the model 
parameters and of the bias conditions. A strong nonlinearity emerges in the 
pre-breakdown region which is also characterized by non-Gaussian noise. 
The results compare well with measurements of electrical breakdown in 
composites and with electromigration experiments in metallic lines.

\end{abstract}
\vspace{1.5cm}
\pacs{Noise; breakdown; percolation; random networks.}
%
%
\begin{multicols} {2}

\section{Introduction}
A large interest exists in the recent literature concerning the
electrical and mechanical stability of different physical systems 
\cite{hermann,bardhan,havlin,ohring,Niemeyer,arcangelis,hansen,sornette92,sahimi92,sornette97,stan_zap_nat,stan_zap,dubson89,chakrabarty,bardhan97,yagil93,heaney,heaney_99,nandi,gingl,sornette96,mukherjee,hirano,pen_prl_fail,pen_physd}. 
In fact, from a technological point of 
view, the increasing level of miniaturization of electronic devices 
enhances the importance of degradation and failure processes \cite{ohring},
\cite{gingl},\cite{pen_physd,scorzoni,vandamme,bloom,vandewalle,houssa,trefan}. On the other hand, from a 
theoretical point of view, the study of the response of disordered systems to 
high external stresses, implied in the study of breakdown and fracture 
processes, is of great help to understand the properties of these systems 
and, more generally, to develop the knowledge of nonequilibrium systems 
\cite{hermann,bardhan,havlin,ohring,Niemeyer,arcangelis,hansen,sornette92,sahimi92,sornette97,stan_zap_nat,stan_zap},\cite{sneppen,stanley,barabasi,pietronero}. 
It is well known that the application of a finite stress (electrical or 
mechanical) to a disordered material generally gives a nonlinear response, 
which ultimately leads to an irreversible breakdown (catastrophic
behavior) in the high stress limit \cite{hermann,bardhan,havlin,ohring}. 
Such breakdown phenomena have been successfully studied by using percolation 
theories 
\cite{hermann,bardhan,havlin},\cite{arcangelis,hansen,sornette92,sahimi92}, 
\cite{dubson89,chakrabarty,bardhan97,yagil93,heaney,heaney_99,nandi,gingl,sornette96,mukherjee,hirano,pen_prl_fail,pen_physd},
\cite{houssa},
\cite{stauffer,sahimi,duxbury87,rammal85,kolek,grier,balberg}.
In particular, by focusing on electrical breakdown, a large attention
has been devoted to the determination, by both theory 
\cite{hermann,bardhan,havlin},\cite{arcangelis,hansen,gingl},
\cite{stauffer,sahimi,duxbury87,rammal85,kolek,grier,balberg} and experiments 
\cite{hermann,bardhan,havlin},
\cite{dubson89,chakrabarty,bardhan97,yagil93,heaney,heaney_99} of the critical 
exponents describing the resistance and the variance of resistance 
fluctuations in terms of the conducting particle (or defect) concentration 
\cite{stauffer,sahimi}. In fact, it is well known that the study of the 
resistance fluctuations is a fundamental tool to extract information about 
the system stability \cite{hermann,bardhan,havlin,ohring},
\cite{vandamme,bloom,vandewalle},\cite{rammal85,kolek,grier,balberg,hooge_dutta,zhang,weissman,jones,nicodemi,icnf01}. 
In spite of the wide literature on the subject few attempts have been made 
so far \cite{sornette96,mukherjee} to describe the behavior of a disordered 
medium over the full range of the applied stress, i.e. by studying the 
response of the system to an external bias when the bias strength covers the 
full range of linear and nonlinear regimes. On the other hand,
important information are expected from this kind of study, like: precursor 
phenomena, role of the disorder, existence of scaling laws, predictability 
of breakdown, etc.

Recently, a new approach has been proposed to investigate the resistance 
noise in thin films over the full range of bias values, from the linear 
regime up to electrical breakdown \cite{pen_prl_stat,pen_physc,pen_pre}. 
In this paper, I will present a brief review of this approach. The
model consists of describing the resistance noise in a conducting film in 
terms of two competing processes taking place in a random resistor network 
(RRN) \cite{pen_prl_stat,pen_physc,pen_pre}. The two processes are driven 
by the joint effect of the electrical bias and of the heat exchange with a 
thermal bath. The breaking of a single resistor of the network is taken as a 
simple and general manner to account for different microscopic mechanisms 
which are responsible for the degradation of the electrical properties of a 
small region of the film. This defect generation process is taken to occur in
competition with an opposite process, named defect recovery, which mimics 
the healing mechanisms. Electromigration of metallic lines 
\cite{ohring,scorzoni} instability of the electrical properties of composites 
or semicontinuous metal films \cite{bardhan,havlin},
\cite{dubson89,chakrabarty,bardhan97,yagil93,heaney,heaney_99,nandi},
\cite{mukherjee,hirano} or soft dielectric 
breakdown of ultra-thin oxides \cite{ohring,vandewalle}, are examples of 
phenomena that can be successfully described by this approach 
\cite{pen_physd,trefan,pen_prl_stat,pen_physc,pen_pre}. In the case of 
electromigration phenomena, for example, the defect generation corresponds 
to the formation of voids induced by the electronic wind, while the defect 
recovery is related to the void healing due to mechanical stress and thermal 
gradients inside a metallic film \cite{ohring,scorzoni}. In the case of 
composites, the electrical instability arises from the opening of new 
conductive channels as a consequence of thermally activated hopping processes 
\cite{bardhan,mukherjee}. On the other hand, the same processes 
can also lead to a suppression of already existing channels. In the case of 
soft dielectric breakdown of ultra-thin oxides it is instead the competition 
between electron trapping and detrapping which plays a key role
on leakage currents \cite{ohring,vandewalle,houssa}. 

All these phenomena and many others \cite{hermann,bardhan,havlin},
\cite{bloom,kish} can be modeled in terms of competition between two opposed 
processes taking place in a RRN. Monte Carlo simulations 
show that, depending on the bias strength, an irreversible failure or a 
stationary state of the RRN can be achieved. By focusing on the steady-state, 
the behavior of the average network resistance and the properties of
the resistance fluctuations are analyzed as a function of the bias. At low 
bias, an effective defect generation probability can be defined controlling 
the network behavior \cite{pen_prl_stat}. In this Ohmic regime, a scaling 
relation has been found between the relative variance of resistance 
fluctuations and the average network resistance \cite{pen_prl_stat}. 
The properties of the nonlinear regime, occurring when the bias overcomes 
a threshold value, are studied for different values of the model parameters 
and for different bias conditions (constant voltage or constant current) 
\cite{pen_physc,pen_pre}. Scaling relations are found relating the average 
resistance and the resistance noise with the external bias. In particular, 
a strong nonlinearity emerges in the pre-breakdown region which is also 
characterized by a non-Gaussian noise. The results compare well with 
measurements of electrical breakdown in composites 
\cite{bardhan,mukherjee,nandi} and in semicontinuous metal films \cite{yagil93}
and with electromigration experiments in metallic 
lines \cite{scorzoni,pen_physd}.

\section{Theory}
The RRN consists of a two-dimensional square-lattice network made of $N_{tot}$
resistors of resistance $r_n$ \cite{stauffer}. Different geometries can be 
studied; here we consider the simplest one which is the square $N \times N$, 
where $N$ determines the linear size of the network. In practical applications
the value of $N$ can be related to the ratio between the size of the film and 
that of the characteristic grain. The RRN exchanges heat with a thermal bath 
at temperature $T_0$ (substrate temperature). A current $I$, kept constant, is 
applied to the network through electrical contacts at the left and right hand 
sides. Alternatively, constant voltage conditions can be considered. The 
current flowing in the n-th resistor of the network is denoted by $i_n$. The
resistances are taken linearly dependent on temperature through a temperature
coefficient of resistance, $\alpha$:
\begin{equation}
r_{n}(T_{n})=r_{0}[ 1 + \alpha (T_{n} - T_0)]
\label{eq:tcr}
\end{equation}
The local temperature $T_n$ is calculated by adopting the biased percolation
model \cite{pen_physica,pen_prl_fail} as:
\begin{equation}
T_{n}=T_{0} + A \Bigl[ r_{n} i_{n}^{2} + {B \over N_{neig}}
\sum_{l=1}^{N_{neig}}  \Bigl( r_{l} i_{l}^2   - r_n i_n^2 \Bigr) \Bigr]
\label{eq:temp}
\end{equation}
where $N_{neig}$ is the number of first neighbors around the n{\em th} 
resistor. The parameter $A$, measured in (K/W), describes the heat coupling 
of each resistor with the thermal bath and it determines the importance of 
Joule heating effects. The parameter $B$ is taken to be equal to $3/4$ to
provide a uniform heating in the perfect network configuration. It must be 
noticed that Eq.~(\ref{eq:temp}) implies an instantaneous thermalization of 
each resistor at the value $T_n$, thus, by adopting Eq.~(\ref{eq:temp}),  
we are neglecting for simplicity time dependent effects which are discussed 
in Ref. \cite{sornette92}.

Here, the initial state of the network is chosen coinciding with the perfect 
network ($r_n \equiv r_0$) of resistance $R_0$. Nevertheless, a disordered 
initial configuration can easily be introduced in the present approach
to analyze the role of the initial disorder on the RRN response. 
We assume that two competing processes act to determine the 
RRN evolution. The first process consists of generating fully insulating 
defects (resistors with very high resistance, i.e. broken resistors) with 
probability \cite{gingl} $W_{D,n} = \exp( -E_D/K_B T_n )$, where $E_D$ is a 
characteristic activation energy and $K_B$ the Boltzmann constant. The second 
process consists of recovering the insulating defects with probability 
$W_{R,n} = \exp (-E_R/K_B T_n)$, where $E_R$ is an activation energy 
characteristic of this second process \cite{pen_physd,pen_prl_stat}. 
Thus, the first process consists in a percolation of broken resistors 
within a network of active resistors. This percolative process is contrasted 
by a recovery process. This second process can also be seen as a 
percolation of active resistors within an insulating network. 
For $A\neq 0$, Eq.~(\ref{eq:temp}) implies that both the processes (defect 
generation and defect recovery) are correlated processes. Indeed, the 
probability of breaking (recovering) a resistor is higher in the so called 
``hot spots'' of the RRN \cite{stauffer}. On the other hand, for $A=0$ 
Eq.~(\ref{eq:temp}) yields $T_n \equiv T_0$, which corresponds to the 
competition of two random processes \cite{stauffer,pen_prl_stat}. 
The same is true for vanishing small bias values, when Joule heating 
effects are negligible.

\begin{figure}[htbp]
\epsfig{figure=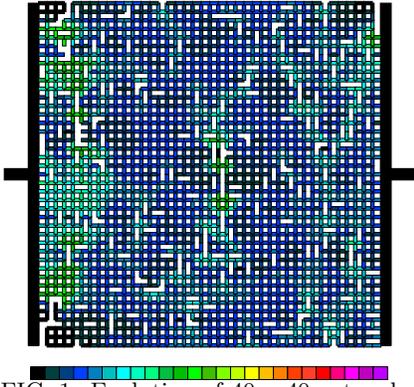,width=5.5cm}
\caption{Evolution of $40 \times 40$ network obtained for a value of the
external current $I=1.2$ (A). The different colors, from black to violet,
correspond to increasing values of $r_n$ from $1$ to $15$ ($\Omega$).
The pattern shown corresponds to the iteration step $t=800$.}
\label{fig:1}
\end{figure}
\begin{figure}[htbp]
\epsfig{figure=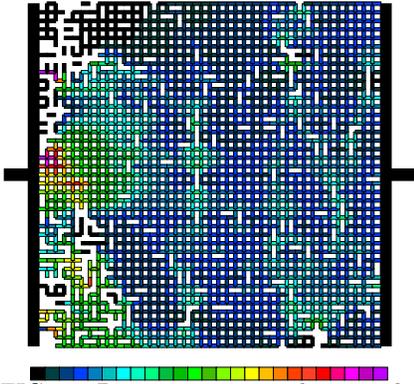,width=5.5cm}
\caption{Pattern corresponding to the same evolution of Fig. 1 and to
the iteration step $t=805$.}
\label{fig:2}
\end{figure}

The sequence, defect-creation and defect-recovery, is then iterated.
Depending on the material parameters ($E_D$, $E_R$, $A$, $\alpha$, $r_0$, $N$)
and on the external conditions (specified by the kind of applied bias 
and by the bath temperature) a steady state or an irreversible breakdown 
characterized by a critical fraction of defects $p_c$ (percolation 
threshold) are reached. In the first case, the network resistance fluctuates 
around an average value $<R>$. In the second case, $R$ diverges due to the 
existence of at least one continuous path of defects between the upper and 
lower sides of the network \cite{stauffer}. We note that in the limit of a 
vanishing bias (two random processes) and infinite lattices 
($N \rightarrow \infty$), the expression:
$E_R < E_D + K_BT_0 \, \ln [1 + \exp(-E_D/K_BT_0)]$ provides a sufficient
condition for the existence of a steady state \cite{pen_prl_stat}.
In the general case of biased processes, no analytical expressions are
available for the steady state condition and thus it is necessary to rely on
numerical simulations only.

\begin{figure}[htbp]
\epsfig{figure=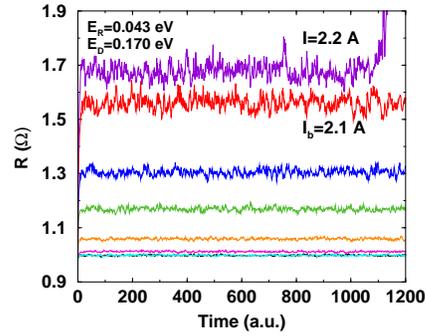,width=5.5cm}
\caption{Resistance evolutions for increasing bias values. Going from 
bottom to top, the curves are obtained for 
$I=0.01, 0.1, 0.5, 1.0, 1.5, 1.8, 2.1$ (A) and they correspond to RRN steady
states. Precisely, the highest curve shows an unsteady evolution of the 
resistance obtained for $I=2.2$ (A), the curve just below corresponds to the 
breakdown current $I_b=2.1$.}
\label{fig:3}
\end{figure}

The evolution of the RRN is obtained by MC simulations carried out according
to the following procedure. (i) Starting from the perfect lattice
with given local currents, the local temperatures $T_n$ are calculated
according to Eq.~(\ref{eq:temp}); (ii) the defects are generated with
probability $W_D$ and  the resistances of the unbroken resistors are changed
as specified by Eq.~(\ref{eq:tcr}); (iii) the currents $i_n$ are calculated
by solving Kirchhoff's loop equations by the Gauss elimination method and the
local temperatures are updated; (iv) the defects are recovered with 
probability $W_R$ and the temperature dependence of unbroken resistors is
again accounted for; (v) $R$, $i_n$ and $T_n$ are finally calculated and the
procedure is iterated from (ii) until one of the two following possibilities
is achieved. In the first, the percolation threshold is reached. In the
second, the RRN attains a steady state; in this case the iteration runs
long enough to allow a fluctuation analysis to be carried out.
Each iteration step can be associated with an elementary time step on an
appropriate time scale (to be calibrated with experiments). In this manner
it is possible to represent the simulated resistance evolution over
either a time or a frequency domain according to convenience.
Except when differently specified, the simulations are performed by taking 
the following values for the parameters, which are chosen as reasonable 
values: $N =75$, $r_0$ ranges from $1 \div 10 \ (\Omega)$, 
$\alpha = 10^{-3}$ (K$^{-1}$), $A=5 \times 10^5$ (K/W), $E_D = 0.170$ (eV), 
$E_R$ in the range $0.026 \div 0.155$ (eV) and $T_0=300$ (K). The values of 
the external bias range from $0.001 \le I \le 3.0$ (A) under constant current 
conditions, and from $0.001 \le V \le 3.5$ (V) under constant voltage 
conditions. Further details about numerical simulations can be found in 
Refs.\cite{pen_prl_fail,pen_prl_stat,pen_physc,pen_pre}.

\begin{figure}[htbp]
\epsfig{figure=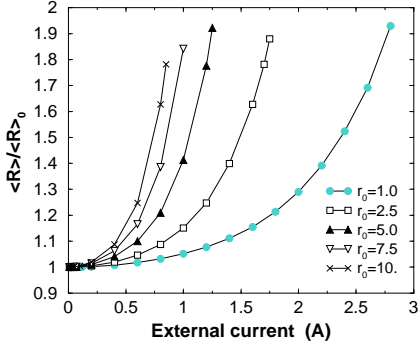,width=5.5cm}
\caption{Average resistance versus the external current. The resistance
is normalized to the linear regime value. The different curves are obtained
for $E_D=0.17$ (eV) and $E_R=0.026$ (eV), while $r_0$ ranges from $1$ to $10$
($\Omega$).}
\label{fig:4}
\end{figure}
\begin{figure}[htbp]
\epsfig{figure=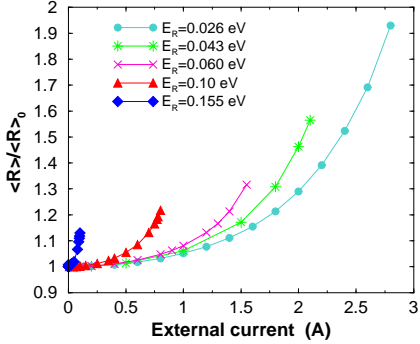,width=5.5cm}
\caption{Average resistance versus the external current. The resistance
is normalized to the linear regime value. The different curves are obtained
for $E_D=0.17$ (eV) and $r_0=1$ ($\Omega$), while $E_R$ ranges from $0.026$
to $0.155$ (eV).}
\label{fig:5}
\end{figure}

\section{Results}
Figures 1 and 2 display the evolution of a $40 \times 40$ network,
made of resistors with $r_0 = 1.0 \ (\Omega)$, characterized by a value of
$E_R = 0.043$ (eV) and subjected to a bias current $I=1.2$ (A). For this
choice of the parameters, the network becomes unstable for currents 
greater than
$I_b=1.1$ (A). Therefore, in the case considered, the network evolves towards 
the electrical breakdown which is reached after 809 iteration steps. Precisely,
Fig. 1 shows the formation of filamented clusters of defects (missing 
resistors) characteristic of biased percolation \cite{pen_prl_fail,kish}. 
Figure 2 reports the pattern very near to the final breakdown. The incipient 
cluster of defects, perpendicular to the direction of the applied current is 
well evident, together with the formation of hot spots. This kind of damage 
pattern perfectly agrees with the pattern observed in metallic lines failed 
as a consequence of electromigration \cite{ohring,scorzoni}.

Typical resistance evolutions of a $75 \times 75$ RRN, with the same
parameters used for Figs. 1 and 2, are reported in Fig. 3 for different
values of the bias current. The first six curves from the bottom correspond to
steady states of the RRN, the upper curve to an evolution toward electrical
breakdown. In particular, it must be noticed that the first two curves (black
and cyan), obtained for $I=0.01$ (A) and $I=0.1$ (A), are practically
overlapping. By contrast, all the other steady state curves fluctuate around 
significantly different average values with increasing fluctuation amplitudes.
Thus, the figure well illustrates a general feature of the model: the ability 
of describing both the linear and the nonlinear regimes up to the breakdown.

\begin{figure}[htbp]
\epsfig{figure=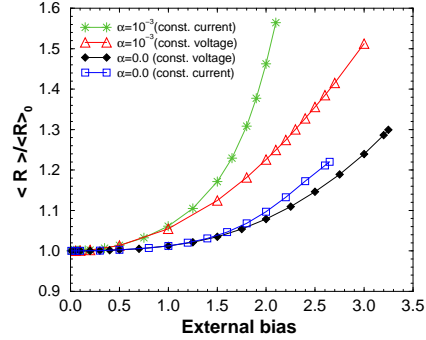,width=5.5cm}
\caption{Normalized average resistance versus the external bias: 
constant current for the green and blue curves, constant voltage for 
the red and black ones. Precisely, the green and red curves are 
obtained for $\alpha=10^{-3}$ (K$^{-1}$) while the blue and black 
for $\alpha=0$. The other parameters are specified in the text.
The bias values are in A and in V units under constant current and 
constant voltage, respectively.}
\label{fig:6}
\end{figure}
\begin{figure}[htbp]
\epsfig{figure=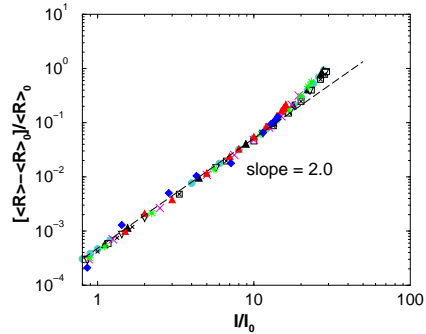,width=5.5cm}
\caption{Log-log plot of the relative variation of the average resistance 
versus the normalized current. Only the nonlinear regime is shown. 
The data correspond to different values of $E_R$ and to different values 
of $r_0$ and they are the same shown in Figs. 4 and 5. The dashed line 
fits the data in the moderate bias region.}
\vspace{-0.3cm}
\label{fig:7}
\end{figure}
\begin{figure}[htbp]
\epsfig{figure=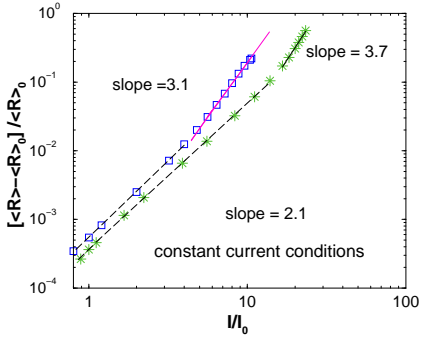,width=5.5cm}
\caption{Log-log plot of the relative variation of the average 
resistance versus the normalized current. Only the nonlinear regime 
is shown. The data indicated with green stars are obtained for 
$\alpha=10^{-3}$ (K$^{-1})$, while the data indicated with blue 
squares for $\alpha=0$. The dashed black lines fit the data with a 
power-law of exponent 2.1, the magenta line with a power-law of 
exponent 3.1 and the black solid line corresponds to an exponent 3.7.}
\label{fig:8}
\end{figure}
\begin{figure}[htbp]
\epsfig{figure=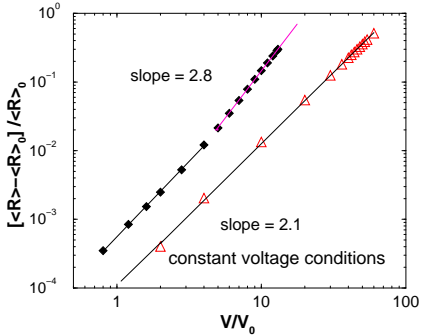,width=5.5cm}
\caption{Log-log plot of the relative variation of the average 
resistance versus the normalized voltage. The data indicated with 
red triangles are obtained for $\alpha=10^{-3}$ (K$^{-1})$, 
while the data indicated with black diamonds for $\alpha=0$. 
The black lines fit the data with a power-law of exponent 2.1 
while the magenta line with a power-law of exponent 2.8.}
\label{fig:9}
\end{figure}

Figure 4 shows the average values of the RRN resistance as a function of the 
bias current. Each value is calculated by considering the time average on a 
single steady-state realization and then averaging over 20 independent 
realizations. Further details can be found in Ref. \cite{pen_pre}. The 
different curves correspond to different values of $r_0$, as specified in the 
figure, and to a common value of $E_R=0.026$ (eV). At the lowest biases the 
average resistance is independent of the bias values and it takes a value 
$<R>_0$, which represents the linear response of the network (Ohmic regime) 
\cite{pen_prl_stat}. For this reason in Fig. 4 the average resistance has been
normalized to $<R>_0$. At increasing biases, when the current overcomes a 
certain value, $I_0$, the average resistance starts to become dependent on the
bias. Thus, $I_0$ sets the current scale value for the onset of nonlinearity. 
The average resistance increases with bias up to a value $<R>_b$, which 
corresponds to a threshold current $I_b$. Above this threshold current the 
RRN undergoes an irreversible breakdown. Details about the criteria used for 
the determinations of $I_0$ and $I_b$ can be found in Ref.\cite{pen_pre}. 
It  must be noticed that by increasing the initial RRN resistance, both $I_b$ 
and $I_0$ decrease but the ratios $I_b/I_0$ and $<R>_b/<R>_0$ are found to be 
independent of the initial resistance value \cite{pen_physc}. Moreover, these
ratios are found independent of the network sizes \cite{unpub}. Both 
these behaviors agree with electrical breakdown measurements performed in 
composites \cite{mukherjee} and in semicontinuous metal films \cite{yagil93}. 

Figure 5 reports the normalized average resistance as a function of the bias 
current for different values of the recovery activation energy, specified in 
the figure. All the curves are now obtained by taking $r_0=1$ ($\Omega$). 
In this case, not only $I_b$ and $I_0$ decrease with $E_R$ but also the 
ratios $I_b/I_0$ and $<R>_b/<R>_0$ diminish \cite{pen_physc}. Thus, in 
contrast to what happens for the effect of the initial resistance,  
the decreasing robustness of the system is associated with a 
reduction of the extent of the nonlinear region. It must be underlined that 
if the value of $E_R$ is sufficiently different from its maximum value 
$E_{R, MAX}$ (determined by the stability condition in the vanishing bias 
limit, discussed in Sec. 2) the threshold current $I_b$ is associated with a 
first order transition. In fact, the ``last'' stable state of the RRN, 
corresponding to $I_b$, is characterized by an average defect fraction 
$<p>_b$ which is smaller than $p_c$. Therefore, the correlation length and 
the correlation time remain finite at the threshold \cite{unpub}, denoting 
a non critical behavior. It must be noticed that the prediction of a first 
order transition is confirmed by experiments \cite{bardhan,mukherjee}.

\begin{figure}[htbp]
\epsfig{figure=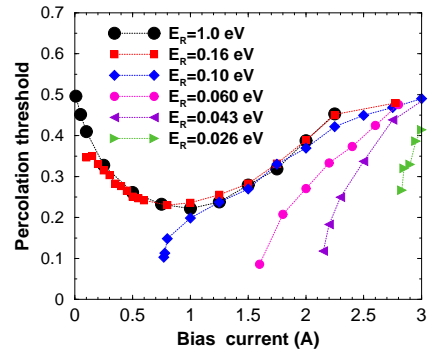,width=5.5cm}
\caption{Percolation threshold versus the external current for different
values of the recovery energy $E_R$ which ranges from $0.026$ (eV) 
to $1.0$ (eV).}
\label{fig:10}
\end{figure}
\begin{figure}[htbp]
\epsfig{figure=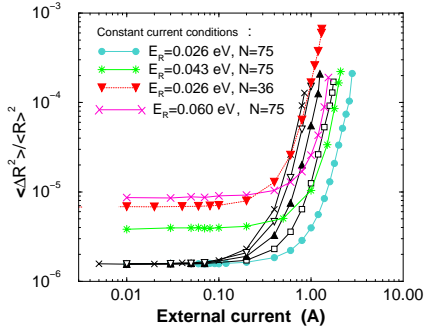,width=5.5cm}
\caption{Relative variance of resistance fluctuations versus the external
current. Magenta, green and cyan curves correspond to decreasing values of
$E_R$. The black curves are obtained for the same value of $E_R$ used for
the cyan curve and they differ for the value of $r_0$. Precisely, cyan
circles $r_0=1.0$, open square $r_0=2.5$, black triangles $r_0=5.0$,
open triangles  $r_0=7.5$, crosses  $r_0=10.0$. The values of $r_0$
are expressed in $\Omega$. The red curve with down triangles is obtained 
for the same parameters of the cyan curve but for a smaller network, 
of size $N=36$.}
\vspace{-0.2cm}
\label{fig:11}
\end{figure}

\begin{figure}[htbp]
\epsfig{figure=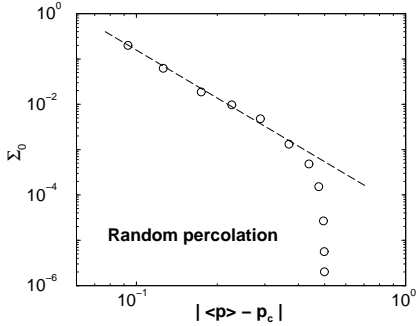,width=5.5cm}
\caption{Relative variance of resistance fluctuations versus $|<p>-p_c|$ 
in the linear regime. This regime corresponds to the competition of two
random processes.}
\vspace{-0.3cm}
\label{fig:12}
\end{figure}

The effect of the temperature coefficient of the resistance, $\alpha$, is 
shown in Fig. 6, where we have also reported the behaviors of the average 
resistance calculated under constant current and under constant 
voltage conditions. The simulations are performed by taking $r_0=1$ and 
$E_R=0.043$ (eV). Similarly to the constant current case, also under 
constant voltage two threshold voltage values exist: $V_0$ and $V_b$,
corresponding to the nonlinearity onset and to the electrical breakdown, 
respectively. The ratio $V_b/V_0$ is found to be significantly higher than 
the ratio $I_b/I_0$ \cite{pen_pre}. This fact reflects the greatest stability 
of the system when biased under constant voltage than under constant 
current. This property is further emphasized by the fact that the increase 
of the resistance in the pre-breakdown region exhibits a lower slope 
under constant voltage than under constant current conditions, as discussed
below. It must be noticed that in spite of the significant difference of the 
ratios $V_b/V_0$ and $I_b/I_0$, the ratio $<R>_b/<R>_0$ remains practically 
the same under the different bias conditions \cite{pen_pre}, in agreement 
with experiments \cite{mukherjee}.

\begin{figure}[htbp]
\epsfig{figure=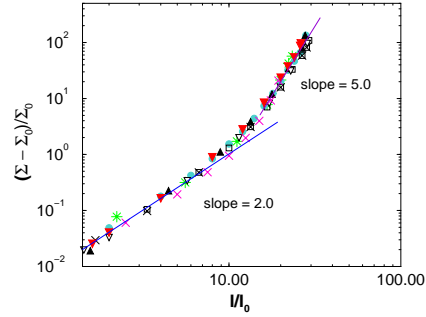,width=5.5cm}
\caption{Log-log plot of the relative variation of $\Sigma$ versus $I/I_0$. 
The data shown are obtained for different values of $E_R$, $r_0$ and $N$ and 
they are the same of Fig. 11. The black line fits the different sets of data 
in the moderate bias region with a power-law of exponent 2.0, while the blue 
line of slope 5.0 in the high bias region is only for a guide to the eyes.}
\vspace{-0.3cm}
\label{fig:13}
\end{figure}

To analyze the dependence on the bias of the average resistance and to 
investigate the existence of scaling relations and their universality 
\cite{stanley}, Fig. 7 reports on a log-log plot the relative variation of 
the average resistance, $\Delta <R>/<R>_0 \equiv (<R>-<R>_0)/<R>_0$, as a 
function of $I/I_0$ for different values of $r_0$ and $E_R$. The data are 
the same of Figs. 4 and 5. The figure shows that all the curves collapse onto 
a single one and that the relative variation of $<R>$ scales with the ratio 
$I/I_0$ as  \cite{pen_physc}:
   \begin{equation}
   \frac{<R>}{<R>_{0}} = g(I/I_{0}), \qquad
   g(I/I_{0}) \simeq 1+ a(I/I_{0})^{\theta}
   \label{eq:biasres}
   \end{equation}
with the exponent $\theta=2.1 \pm 0.1$ and $a$ a dimensionless coefficient. 
However, it must be noticed that a superquadratic behavior of $<R>$ emerges in
the pre-breakdown region, where the relative variation of $<R>$ is 
characterized by a power law $(I/I_0)^{\theta_I}$ with an exponent 
$\theta_I=3.7 \pm 0.3$ \cite{pen_pre}. Furthermore, this superquadratic 
behavior emerges only for RRN sufficiently robust. In fact, at increasing the 
recovery activation energy $E_R$, the ratio $I_b/I_0$ becomes smaller, i.e. 
the stability region is reduced and the dependence on the current of the 
average resistance can remain quadratic over the entire region.

\begin{figure}[htbp]
\epsfig{figure=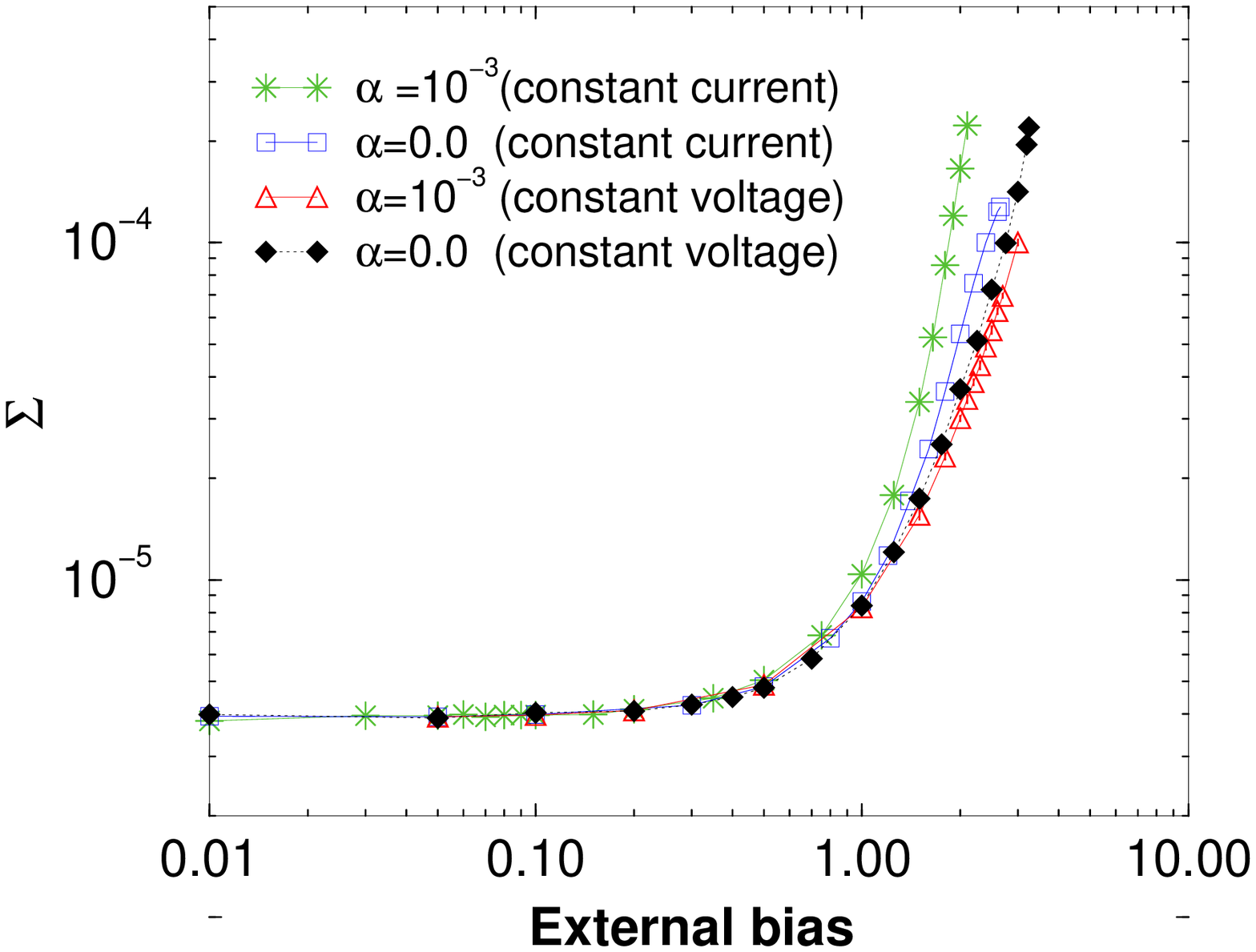,width=5.5cm}
\caption{Relative variance of resistance fluctuations versus the external
bias: constant current for the green and blue curves, constant voltage for
the red and black ones. Precisely, the green and red curves are obtained for
$\alpha=10^-{3}$ (K$^{-1}$) while the blue and black for $\alpha=0$. The 
other parameters are specified in the text. The bias values are in A and in 
V units under constant current and constant voltage, respectively}.
\vspace{-0.5cm}
\label{fig:14}
\end{figure}
\begin{figure}[htbp]
\epsfig{figure=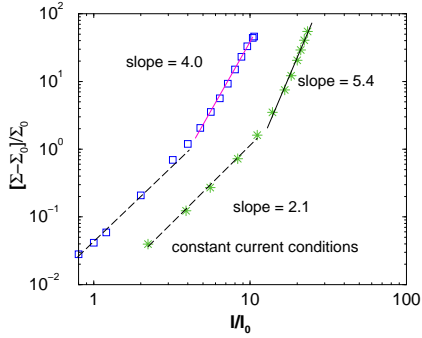,width=5.5cm}
\caption{Log-log plot of the relative variance of resistance fluctuations
versus the normalized current. The data indicated with green stars are
obtained for $\alpha=10^-{3}$ (K$^{-1})$, while the data indicated with blue
squares for $\alpha=0$. The black dashed lines fit the data with
a power-law of exponent 2.1, the black solid line with a power-law of
exponent 5.4, while the magenta solid line corresponds to an exponent 4.0.}
\vspace{-0.3cm}
\label{fig:15}
\end{figure}
\begin{figure}[htbp]
\epsfig{figure=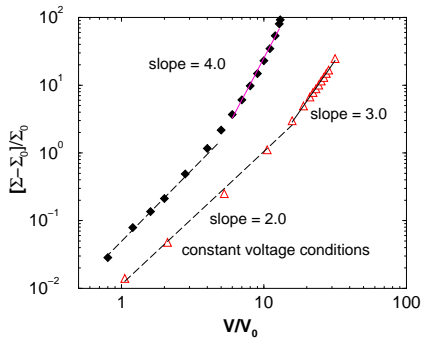,width=5.5cm}
\caption{Log-log plot of the relative variance of resistance fluctuations
versus the normalized voltage. The data indicated with red triangles are
obtained for $\alpha=10^-{3}$ (K$^{-1})$, while the data indicated with black
diamonds for $\alpha=0$. The black dashed lines fit the data with
a power-law of exponent 2.0, the black solid line with a power-law of
exponent 3.0, while the magenta solid line corresponds to an exponent 4.0.}
\vspace{-0.1cm}
\label{fig:16}
\end{figure}
\begin{figure}[hbtp]
\epsfig{figure=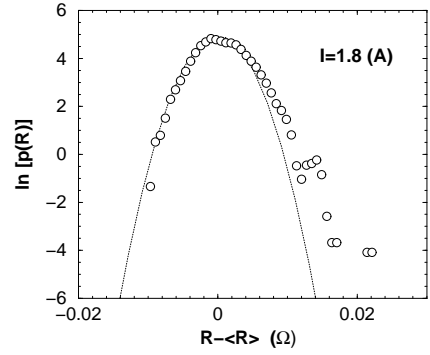,width=5.5cm}
\caption{Distribution function of the resistance fluctuations for an
applied current of $I=1.8$ (A), the other parameters are specified in the
text. The scale is a linear-log, therefore the dashed curve corresponds
to a Gaussian distribution.}
\label{fig:17}
\end{figure}
\begin{figure}[hbtp]
\epsfig{figure=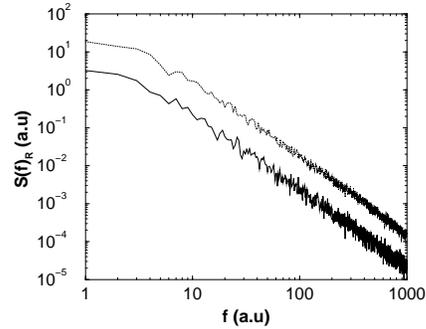,width=5.5cm}
\caption{Power spectral density of resistance fluctuations under constant
current conditions. The solid curve is obtained for $I=1.5$ (A), the dotted
one for $I=1.8$ (A).}
\vspace{-0.3cm}
\label{fig:18}
\end{figure}

In the log-log plot reported in Fig. 8 we compare the relative variations 
of the resistance versus the ratio $I/I_0$ calculated for two different 
values of the temperature coefficient: $\alpha=10^{-3}$ (K$^{-1})$ and 
$\alpha=0$ (the data are the same as those shown in Fig. 6 with stars and 
open squares respectively). We can see that the two sets of data do not 
collapse onto the same curve, i.e. they do not belong to the same universality
class. Moreover, the effect of a nonzero value of $\alpha$ is to significantly 
enhance the superquadratic dependence of the resistance on the bias under 
constant current conditions. By contrast, the opposite is true under constant 
voltage conditions, as shown in Fig. 9, where the data obtained for 
$\alpha=10^{-3}$ (K$^{-1})$ are again compared with those obtained for 
$\alpha=0$ (same data shown in Fig. 6 with triangles and full 
squares, respectively). In fact, in this case, the effect of $\alpha$ is 
to depress the superquadratic behavior. Remarkably, under constant 
voltage conditions, the average resistance scales quadratically with the 
applied voltage over the whole region of bias values up to breakdown, 
as \cite{pen_pre}: 
\begin{equation}
{<R>_V \over <R>_0} = 1 + a' \left ({V \over V_0} \right )^{\theta}
\label{eq:res_v}
\end{equation}
where $a'$ is a dimensionless coefficient and the value of $\theta$ is the 
same of that in Eq.~(\ref{eq:biasres}).

The quadratic dependence on the bias in the moderate bias region, common to 
all the data shown in Figs. 7, 8 and 9, must be considered a general 
feature of the model, in agreements with experiments \cite{mukherjee} and 
can be explained as follows. It is well known \cite{stauffer} that the 
resistance of sufficiently large RRNs subjected to random percolation is 
related to the fraction of broken resistors by the scaling relation: 
\begin{equation}
R \sim | p - p_{c} |^{-\mu} 
\label{eq:res_scal}
\end{equation}
where $p_c \equiv p_{c0}=0.5$ for a square lattice and $\mu$ takes a 
universal value, known from very accurate calculations: $\mu=1.303$ for 
two-dimensional RRNs \cite{stauffer}. This scaling relation should hold also 
for steady states of RRNs resulting from the superposition of two opposite 
random processes. In this case, Eq.~(\ref{eq:res_scal}) relates the average 
network resistance $<R>$ with the average fraction of defects $<p>$. 
Numerical simulations reported in Ref. \cite{pen_prl_stat} confirm this 
statement. On the other hand, when a single biased percolation is present 
$p_c$ becomes a function of the bias strength \cite{pen_mcs,pen_physica} 
(see black circles in Fig. 10), moreover the value of $\mu$ is no more 
universal and it depends on the biasing conditions 
\cite{pen_physica,pen_prl_fail}. When two competing biased 
processes are present, leading to steady states of the RRN, the use of 
Eq.~(\ref{eq:res_scal}) becomes quite problematic. In fact, in this case 
$p_c$ depends also on the recovery activation energy, as shown in Fig. 10. 
In this figure the black circles represent $p_c$ versus the bias when 
$E_R \gg E_D$ so that no recovery of defects occurs. In this case the RRN 
is unstable for any value of the applied current and, at vanishing bias, 
$p_c$ takes the random percolation value $p_{c0}$. At increasing bias, 
the correlated growth of defects along filamented clusters reduces $p_c$. 
At still higher values of the bias, a multi-channel filamentation emerges 
\cite{pen_mcs} which makes $p_c$ an increasing function of $I$. When 
$E_R < E_D$ a stability region appears for $I < I_b$, whose extent increases 
at decreasing $E_R$, as shown in Fig. 5. Therefore, the corresponding curves 
in Fig. 10 start at progressively higher values of $I$. In spite of these 
complications, at moderate bias and when $<p> \ll p_c$, by truncating 
Eq.~(\ref{eq:res_scal}) to the first order in  $<p>/p_c$, we can write: 
$\Delta <R> \approx C(\mu_0,p_{c0}) \ \Delta<p>$. Furthermore, in the 
spirit of a mean-field theory, it is possible to see \cite{pen_pre} that 
the relative variation of the average defect fraction: 
$\Delta<p>/<p>_0 \ \propto \ (I/I_0)^2$. Thus, also 
$\Delta <R> \ \propto \ (I/I_0)^2$. At high bias, terms of order higher than 
the first in the expansion of $<R>$ in terms of $<p>/p_c$ and the dependence 
on the bias of $p_c$ become important. Furthermore, $\Delta <p>$ is no more a
quadratic function of the bias \cite{pen_pre}. Therefore, the average 
resistance acquires the superquadratic behavior shown in Figs. 7, 8 and 9.
On the basis of the previous results and of the above discussion, the 
behavior of the average resistance in the moderate bias region is expected 
to be a universal feature. Of course a final validation of this statement 
would require a study of the effect of the network topology (non square 
lattices, etc.) and of the boundary conditions.

The relative variance of resistance fluctuations, 
$\Sigma \equiv <\Delta R^2>/<R>^2$, is reported in Fig. 11 as a function of 
$I$ and for different $r_0$ and $E_R$ values. The data correspond to the same 
simulations shown in Figs. 4 and 5 and they are obtained by using the same 
procedure of time averaging over a single simulation and then ensemble 
averaging over $20$ realizations. Only the three lowest values of $E_R$ 
used in Fig. 5 have been reported in Fig. 11 and with the same symbols. 
In addition, the red curve with down triangles shows the relative variance 
$\Sigma$ obtained for a network of size $36 \times 36$. All the other 
parameters are the same as those ones used for the curve with cyan circles 
($E_R=0.026$ eV and $r_0=1. \Omega$). At low bias, the relative variance of 
resistance fluctuations is found to achieve a constant value $\Sigma_0$ which 
represents an intrinsic property of the system. On the other hand, a strong 
increase of $\Sigma$ characterizes the nonlinear regime occurring for 
$I >I_0$.

By focusing on $\Sigma_0$, we can see in Fig. 11 that this quantity is 
independent of $r_0$ but strongly dependent on the recovery activation energy 
$E_R$. In fact $\Sigma_0$, corresponding to the vanishing bias limit, arises 
from the competition of two random processes. The study of the resistance 
noise in RRNs subjected to two random processes, has been performed in 
Ref. \cite{pen_prl_stat} and it pointed out the following results. $\Sigma_0$ 
exhibits two regimes as a function of $|<p> - \ p_{c0} \ |$ (see Fig. 12). 
A nearly perfect network regime occurs when $<\Delta R^2>_0$ is less 
than $R_0^2/2N^2$, while a disordered network regime appears in the opposite 
case. In the first case, the resistance noise is directly proportional to the 
fraction of defects \cite{pen_prl_stat}. In the other case (disordered 
network), the breaking and the recovering of the backbone resistors 
\cite{stauffer} result in an enhancement of the resistance noise. It is 
noteworthy that the first regime is a finite size effect which disappears 
in the limit of an infinitely large network \cite{pen_prl_stat}. 
In the second regime the data follow closely the scaling relation:
\begin{equation}
\Sigma_0 \sim |<p> - \ p_{c0} \ |^{-k}
\label{eq:ran_pscal}
\end{equation}
with $k=3.1$ \cite{pen_prl_stat}. A similar scaling relation, with a 
different critical exponent $k_f=1.12$, has been found between the relative 
variance of resistance fluctuations and $|<p> - \ p_{c0}|$ in the case of 
1/f noise in random networks \cite{rammal85}. On this respect, it must be 
noticed that the expression 1/f noise is used as a shorthand for a noise 
(i) spatially uncorrelated, (ii) statistically uniform and (iii) sufficiently 
small on each elementary resistors of the network to be treated only to the 
lowest order. Thus, it is this last feature (iii) which makes the crucial 
difference between the noise in Ref. \cite{rammal85} and the $\Sigma_0$ noise
given by Eq.~(\ref{eq:ran_pscal}), which instead arises from the random 
switch-off and switch-on of the elementary resistors. As a consequence, 
the noise in Eq.~(\ref{eq:ran_pscal}) is strongly sensitive to the average 
defect fraction. 
By combining $\Sigma_0 \sim |<p> - \ p_{c0} \ |^{-k}$ with 
$<R>_0 \sim |<p> - \ p_{c0} \ |^{-\mu}$, one obtains in the disordered 
network regime: 
\begin{equation}
\Sigma_0 \sim <R>_0^s
\label{eq:ran_rscal}
\end{equation}
with $s = k/\mu = 2.6$. A more detailed analysis of the properties of 
$\Sigma_0$ can be found in Ref. \cite{pen_prl_stat}. 

Now, by considering the behavior of $\Sigma$ in the nonlinear regime, it is 
convenient to report its relative variation, $(\Sigma -\Sigma_0)/\Sigma_0$,
on a log-log plot as a function on $I/I_0$. This is done in Fig. 13, where 
the same data of Fig. 11 are shown. Similarly to what happens for the 
average resistance, two regions can be identified in the nonlinear regime. 
At moderate bias, $\Sigma$ scales quadratically with the current, as:
        \begin{equation}
        \frac{\Sigma}{\Sigma_{0}} = f(I/I_{0}), \qquad
        f(I/I_{0}) \simeq 1 + c (I/I_{0})^{\eta}
        \label{eq:nois}
        \end{equation}
with $\eta=2.0 \pm 0.1$ and $c$ a dimensionless coefficient. At higher 
biases, in the pre-breakdown region, a strong superquadratic dependence 
appears \cite{pen_physc,pen_pre}. It must be remarked that the behavior of 
$\Sigma$ in the pre-breakdown region is found to be dependent on $E_R$. 
In fact, in this region:
\begin{equation}
{\Sigma_I \over \Sigma_{0}} \sim \left ({I \over I_0} \right )^{\eta_I}
\label{eq:sigm_ib}
\end{equation}
with $4.5 \leq \eta_I \leq 5.5$, depending on $E_R$. Therefore, the 
straight-line with slope 5 in Fig. 13 has been drawn for a guide to the eyes. 
This behavior of the relative variance of resistance fluctuations in the 
pre-breakdown region is different from that obtained in the same region 
for the average resistance. In fact, as already discussed in connection
with Fig. 7, $\theta_I$ is found to be practically independent of the 
$E_R$ value. The strong nonlinearity and nonuniversality of $\Sigma$ in the 
pre-breakdown region has been observed in electrical noise measurements in 
conducting polymers \cite{nandi} and in other materials 
\cite{bardhan,vandamme,weissman,bloom}. 

The effect on $\Sigma$ of $\alpha$ and of the different bias conditions 
(constant current or constant voltage) is considered in Fig. 14. The values 
of $\Sigma$ in these figures are obtained from simulations performed 
for $E_R=0.043$ (eV) and $r_0=1$ ($\Omega$) (as in Figs. 6, 8 and 9). 
The data indicated by stars and open squares are obtained under constant 
current conditions with $\alpha=10^{-3}$ (K$^{-1})$ and $\alpha=0$
respectively, while data denoted with triangles and black diamonds are 
obtained under constant voltage conditions for the two different values of 
$\alpha$. Fig. 14 shows that the linear regime value, $\Sigma_0$, is 
independent of $\alpha$ and bias conditions. By contrast, a 
significantly different behavior of $\Sigma$ is found in the nonlinear regime.
In particular, the increase of $\Sigma$ in the pre-breakdown region is 
steeper under constant current than under constant voltage conditions. 
This feature is ascribed to the better stability of the RRN under constant 
voltage, since resistance fluctuations in excess over the mean value are 
damped in this condition while they are enhanced under constant current 
conditions. 
The log-log plots of the relative variation of $\Sigma$ as a function of 
$I/I_0$ and $V/V_0$, reported in Figs. 15 and 16 respectively, 
display the following features. The quadratic dependence of $\Sigma$ on 
the bias in the moderate bias region is a feature independent of the 
values of $\alpha$ and of the bias conditions. Further investigations,
concerning the effect of the network structure and of the geometry of the 
electrical contacts on this feature are necessary to establish the 
universality of this behavior. By contrast, the superquadratic
behavior in the pre-breakdown region is strongly influenced by both the 
$\alpha$ value and the bias conditions. Precisely, the simulations in Fig. 15, 
performed under constant current conditions with $\alpha=0$ 
and with $\alpha=10^{-3}$ (K$^{-1}$), show a strong enhancement of the 
relative variance of resistance fluctuations induced at high bias 
by a positive value of the temperature coefficient of the resistance.
On the other hand, by comparing the superquadratic behavior obtained, with
$\alpha=10^{-3}$ (K$^{-1}$), under constant current conditions and under
constant voltage, we can see that the value of $\eta_I$ is drastically 
reduced from $5.4$ to $3.0$ under constant voltage. An extended discussion 
of these behaviors can be find in Ref. \cite{pen_pre}. 

The Gaussian properties of the fluctuation amplitudes has also been 
investigated \cite{pen_pre}. Figure 17 reports the distribution function of 
the resistance fluctuations $p(R)$ obtained for a bias current $I=1.8$ (A) 
(pre-breakdown region). The values of the parameters are the same used 
for the curves in Fig. 3. The parabola in Fig. 17 represents 
the fit with a Gaussian distribution. The figure shows a strong enhancement 
of the probability for the positive fluctuations with respect to the Gaussian 
distribution. This non-Gaussianity of the fluctuation amplitudes increases 
at increasing current or at increasing temperature \cite{pen_pre}, when 
approaching the breakdown conditions. This emergence of a non-Gaussian 
behavior near the breakdown can be considered a relevant precursor of 
failure. Experiments performed in different materials near the electrical 
breakdown confirm this non-Gaussianity of fluctuations 
\cite{vandewalle,weissman,hardner,seidler}.

Finally, Fig. 18 displays the spectral densities of resistance fluctuations 
obtained for two increasing values of the bias current: $I=1.5$ and 
$I=1.8$ (A). The parameters are the same as those in Fig. 17. 
The spectral densities have been calculated by Fourier transforming the 
corresponding correlation functions $C_{\delta R}(t)$. The spectra are 
found to be Lorentzian in agreement with experiments 
\cite{ohring,scorzoni,vandamme,weissman}. Moreover, the analysis of the 
correlation functions for different bias values and for different substrate 
temperatures shows that the correlation time is only weakly dependent on the 
bias while it is strongly dependent on the temperature \cite{unpub}. 

\vspace{-0.5cm}
\section{Conclusions}
I have presented a short review of a recently developed 
\cite{pen_physc,pen_prl_stat,pen_pre} approach which allows the study of the 
resistance noise over the full range of bias values, from the linear regime 
up to electrical breakdown. Resistance noise is described in terms of two 
competing processes taking place in RRNs. The two processes consist of 
the breaking and recovering of the elementary network resistors. The 
probabilities of the two processes are controlled by an electrical bias and 
by the external temperature. Monte Carlo simulations are performed to 
investigate the RRN behavior as a function of the bias; steady state
or electrical failure have been found. At the lowest biases, the two 
processes are practically random and an effective defect generation 
probability can be defined which controls the network behavior. In this Ohmic 
regime, a scaling relation has been found between the relative variance of 
resistance fluctuations and the average resistance \cite{pen_prl_stat}. 
The value of the critical exponent is significantly higher than that 
associated with 1/f noise. The properties of the nonlinear regime, occurring 
when the bias overcomes a threshold value, are studied for different values 
of the material dependent parameters and for different bias conditions 
(constant voltage or constant current) \cite{pen_physc,pen_pre}. 
In general, two regions can be identified in the nonlinear regime: 
a moderate bias region, where both the average resistance and the relative 
variance of resistance fluctuations scale quadratically with the bias, and a 
pre-breakdown region where these quantities exhibit a superquadratic 
dependence. The quadratic behavior in the moderate bias region has been found 
independent of the values of the model parameters and of the bias conditions. 
A final conclusion about the universality of this behavior requires 
the investigation of the role of the network topology and of the electrical 
contact geometry. Remarkably, under constant voltage conditions, 
it has been found that the average resistance scales quadratically over 
the full range of voltage values up to breakdown. Moreover, it must be 
underlined the strong effects on the pre-breakdown properties of the 
temperature coefficient of the resistance. Furthermore, non-Gaussian noise 
has been found in the pre-breakdown region under both bias conditions.
The results obtained by simulations are found to agree satisfactorily 
with electrical and noise measurements performed in composites 
\cite{bardhan,mukherjee,nandi} and in semicontinuous metal films
\cite{yagil93}, and in the degradation processes of metallic lines due to 
electromigration \cite{scorzoni,pen_physd}.

\section*{Acknowledgments}
I'm grateful to Dr. Z. Gingl and Prof. L. B. Kish who introduced me to
the biased percolation approach of failure problems. I also thank Prof. L. 
Reggiani, Drs. G. Trefan and E. Alfinito who collaborated to these results. 
Finally I thank INFM for financial support through the STATE project 
and ASI for support through the project I/R/056/01.

%
%

\end{multicols}

\end{document}